\documentclass{bmvc2k}

\usepackage{kotex}
\usepackage{comment}
\usepackage{adjustbox}

\title{DALE : Dark Region-Aware Low-light Image Enhancement}

\addauthor{Dokyeong Kwon$^*$}{dkdk6638@gmail.com}{1}
\addauthor{Guisik Kim$^*$}{specialre@naver.com}{1}
\addauthor{Junseok Kwon$^\dagger$}{jskwon@cau.ac.kr}{2}
\footnotetext{ Indicates equal contributions. $^\dagger$ Corresponding author.} 
\addinstitution{
 School of\\
 Computer Science and Engineering\\
 Chung-Ang University\\
 Seoul, Korea
}
\runninghead{KWON, KIM, KWON}{Dark Region-Aware Low-light Image Enhancement}

\newcommand{\etal}{\textit{et al}. }
\newcommand{\ie}{\textit{i}.\textit{e}., }
\newcommand{\eg}{\textit{e}.\textit{g}., }
\begin{document}
\maketitle
\
\begin{abstract}
In this paper, we present a novel low-light image enhancement method called dark region-aware low-light image enhancement (DALE), where dark regions are accurately recognized by the proposed visual attention module and their brightness are intensively enhanced.
Our method can estimate the visual attention in an efficient manner using super-pixels without any complicated process. 
Thus, the method can preserve the color, tone, and brightness of original images and prevents normally illuminated areas of the images from being saturated and distorted. 
Experimental results show that our method accurately identifies dark regions via the proposed visual attention, and qualitatively and quantitatively outperforms state-of-the-art methods.
\end{abstract}

\section{Introduction}
\label{sec:intro}
Real-world images for outdoor scenes typically contain low-light areas, especially if the images are captured during nighttime or there exists backlit.
However, using these low-light images, conventional computer vision algorithms (\eg object detection and tracking) cannot produce accurate results, because low-light regions cause images to lose local details and significantly reduce image quality.  
Therefore, low-light image enhancement is essential to prevent conventional computer vision algorithms from degrading their performance.

Low-light image enhancement has a long history. 
For example, Pizer \etal~\cite{pizer1987adaptive} enhanced the brightness and contrast of images based on histogram equalization~\cite{Zuiderveld:1994:CLA:180895.180940}. 
The methods in~\cite{kimmel2003variational, shen2009color} introduced retinex theory~\cite{land1971lightness} for low-light enhancement and used illumination information. 
High dynamic range (HDR)-based methods~\cite{ma2015multi, fu2016fusion} have been proposed to enhance the brightness of images, where HDR requires to combine multiple images with different exposures for the same scene. 
\begin{figure*}[t]
    \begin{minipage}[b]{1.0\linewidth}
        \centering
        \includegraphics[width=0.99\linewidth]{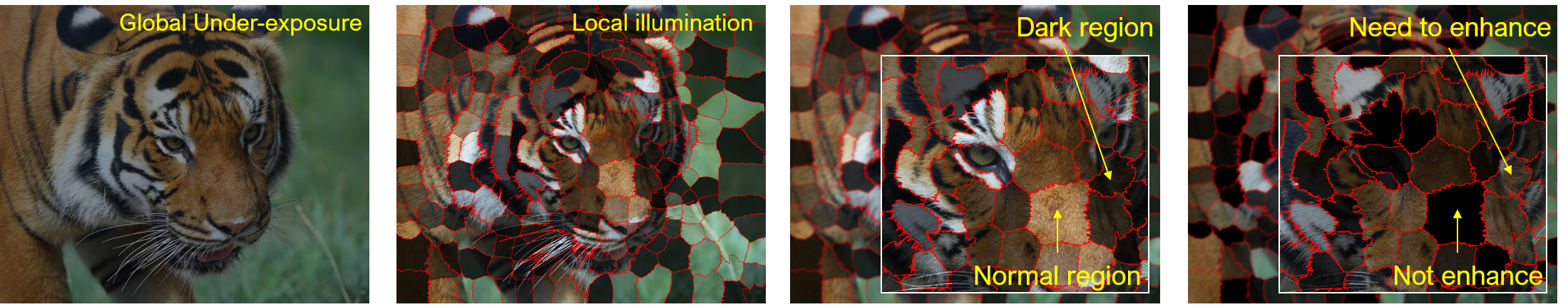}
    \end{minipage}
     \begin{minipage}{0.24\linewidth}
       \centering
        {(a) Original image }
    \end{minipage}   
    \begin{minipage}{0.48\linewidth}
       \centering
        {(b) Local illumination synthesis }
    \end{minipage}  
        \begin{minipage}{0.24\linewidth}
       \centering
        {(c) Dark-aware visual attention}
    \end{minipage}  
  \caption{\textbf{Example of the proposed visual attention for low-light enhancement.} 
  (b) Our method applies a different level of local illumination to each superpixel of a given image. 
  (c) Our method can generate the attention map, where dark areas that need to be enhanced have large values and bright areas have small values.}
 \label{fig:superpixel_discription}
     \vspace{-2mm}
\end{figure*}

Recently, several methods successfully implemented deep neural networks to solve low-level computer vision problems (\eg image super-resolution~\cite{EDSR}, dehazing~\cite{cai2016dehazenet}, and deraining~\cite{ren2019progressive}).
Low-light image enhancement problems have been also addressed in deep learning frameworks~\cite{Kim_ICIP19}.
However, it is nontrivial to train deep neural networks for low-level vision problems, because the networks require be a large number of paired data (\ie image and ground truth) for the training.
To collect a large number of paired data, in super-resolution problems, low-resolution images are synthetically generated by applying noise and blur to high-resolution images.
Low-light image enhancement methods~\cite{Chongyi18, Shen17} also synthetically generate paired data by applying several different levels of global illumination to the images. 
The methods using globally illuminated training images can improve the overall brightness and contrast of images.
However, they cannot accurately enhance the brightness of locally illuminated regions, which are frequently contained in real-world images. 
Applying global illumination is more problematic, if both bright and dark areas exist simultaneously in an image; bright areas of original images can be over-saturated by global illumination. 
To solve this problem, methods in~\cite{lv2019attention, jiang2019enlightengan, lv2018mbllen} proposed to synthesize realistic illumination. 
EnlightenGAN~\cite{jiang2019enlightengan} performed low-light image enhancement in an unsupervised manner without paired data. 
Methods in~\cite{jiang2019enlightengan, lv2019attention, atoum2019color} employed attention modules for low-level vision tasks to focus on the areas, which need to be improved. 
However, in low-light image enhancement, the attention mechanism has not been actively studied and was used only to estimate the illumination channel  as an attention map~\cite{jiang2019enlightengan, Chen2018Retinex}. 

In this paper, we construct a new dataset, which can be used to learn a visual attention map.
Then, the proposed method enhances the brightness of dark areas, which can be recognized by the aforementioned attention map. 
We synthesize differently illuminated superpixels and generate a locally illuminated image dataset, as shown in Fig.\ref{fig:superpixel_discription}(b).
As a result, our method can produce more accurate low-light enhanced images than existing deep-learning based methods. 
Fig.\ref{fig:superpixel_discription}(c) shows an estimated visual attention map, where dark areas are accurately recognized and described using large values.  

The contributions of our method is as follows: 
    \vspace{-3mm}
\begin{itemize}
    \item We present a new attention module to recognize dark areas. 
    For this purpose, we synthesize images to train the attention map, where each superpixel of the images has a different local illumination. 
    Experimental results demonstrate the effectiveness of the proposed dark-aware visual attention. 
    \vspace{-2mm}
    \item We propose a novel low-light enhancement method using the proposed dark-aware visual attention. 
    We call this method dark region-aware low-light image enhancement (DALE). 
    Our method can intensively enhance the brightness of dark areas, while preserving the brightness of other areas. 
    \vspace{-2mm}
    \item We exhaustively conduct the experiments to demonstrate the effectiveness of the proposed method and provide a locally illuminated image dataset, which is used in all experiments.
    This dataset will be publicly available to re-train conventional low-light enhancement methods and improve their accuracy.    
\end{itemize}

\section{Related Work} \label{sec:intro}

\subsection{Low-light image enhancement}
Non-deep learning-based techniques such as histogram equalization and its variations~\cite{pizer1987adaptive} have been commonly adopted for low-light enhancement, in which contrast-limited adaptive histogram equalization~\cite{Zuiderveld:1994:CLA:180895.180940} showed the best performance. 
Retinex theory was employed for low-light enhancement in the single-scale retinex-based method~\cite{land1971lightness}.
Multi-scale retinex with color restoration~\cite{jobson1997multiscale} enhanced the single-scale retinex method by adding the color restoration process. The haze-model was employed to solve low-light enhancement problems \cite{Dong}, while low-light images are similar to reversed hazy images. Lime~\cite{Guo17} used a structure prior to enhance the brightness of low-light regions and outperformed naturalness preserved enhancement~\cite{Naturalness}, multi-deviation fusion~\cite{fu2016fusion}, and simultaneous reflection and illumination estimation~\cite{Xueyang}. 
LECARM~\cite{Ren2019LECARMLI} presented a new camera response model based on the exposure ratio estimation for each pixel. 
Inspired by the human visual system, BIMEF~\cite{Zhenqiang} proposed a multi-exposure fusion technique for low-light enhancement. 
Deep learning methods, MSRNet~\cite{Shen17} and LLNet~\cite{lore2017llnet}, proposed stacked sparse denoising autoencoders to solve low-light enhancement problems. 
LightenNet \cite{Chongyi18} estimated the illumination map using deep neural networks to improve the brightness of images, while RetinexNet~\cite{Chen2018Retinex} estimated both reflection and illumination maps, inducing further improvement.
DeepFuse~\cite{prabhakar2017deepfuse} solved low-light enhancement problems by combining HDR with an unsupervised learning method. MBLLEN~\cite{lv2018mbllen} improves performance using multi-branch fusion methods. 
Lv and Lu \cite{lv2019attention} improved the accuracy of low-light enhancement using attention mechanisms and produced the illumination map.
In contrast, our method directly recognizes dark areas using visual attention modules.  
Thus, our method more intensively enhances the brightness of dark areas. 

Most deep-learning-based methods have a difficult in gathering sufficient training images. 
To address this problem, methods~in \cite{Chongyi18, Shen17} synthesized training data by applying global illumination to the entire image. 
However, these methods are not suitable for real-world images, which contain both low-light and normal illumination areas concurrently. 
Recently, low-lightGAN~\cite{Kim_ICIP19} proposed a generative adversarial network (GAN) in low-light enhancement. 
EnlgithenGAN~\cite{jiang2019enlightengan} present a low-light enhancement method, which can be trained without paired dataset in an unsupervised manner. In contrast to these methods, our method synthesizes training data using superpixels efficiently.

\subsection{Attention Mechanism}
Attention mechanisms have been widely used in many of recent computer vision tasks. 
This attention mechanism stem from the human perception system, where the human brain typically focus on important areas. 
In deep learning, the attention mechanism has been implemented using convolution layers, which results in channel and spatial attention networks. 
There exists a different type of attention mechanisms, which focuses on an image itself instead of convolution layers.
For example, in deraining problems, the method in~\cite{qian2018attentive} recognized raindrop areas using the attention mechanism.

In low-light image enhancement problems, attention-guided method in~\cite{lv2019attention} used the attention map and EnlightenGAN~\cite{jiang2019enlightengan} proposed a self-regularized attention map to enable unsupervised learning.
In contrast to the aforementioned methods that estimate the attention maps to obtain exposure and illumination information, our method directly uses the attention map to enhance the brightness of dark areas.

\begin{figure*}[t]
    \begin{minipage}[b]{1.0\linewidth}
        \centering
        \includegraphics[width=0.99\linewidth]{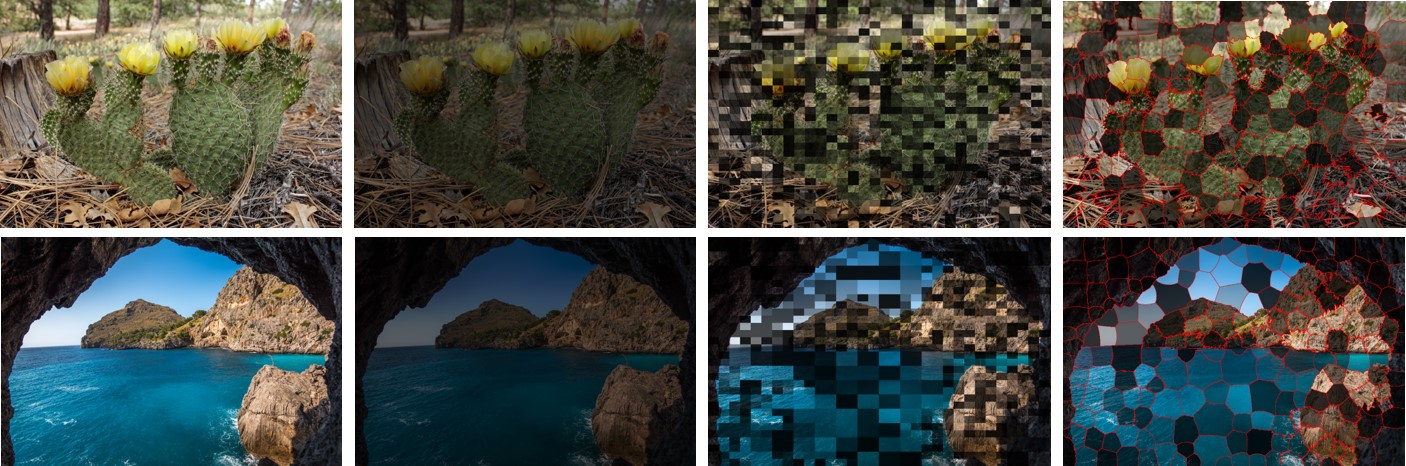}
    \end{minipage}
         \begin{minipage}[b]{0.24\linewidth}
            \centering
            {(a) }
    \end{minipage}
           \begin{minipage}[b]{0.24\linewidth}
            \centering
            {(b) }
    \end{minipage}
            \begin{minipage}[b]{0.24\linewidth}
            \centering
            {(c) }
    \end{minipage}
            \begin{minipage}[b]{0.24\linewidth}
            \centering
            {(d) }
    \end{minipage}
  \caption{\textbf{Proposed local illumination method for training data synthesis.} (a) Input image, (b) Global illumination method, (c) Quad-tree-based method~\cite{Kim_ICIP19}, and (d) Our superpixel-based method.}
 \label{fig:illumination synthesis compare}

\end{figure*}

\section{The Proposed Method}
We construct a new low-light driven training dataset (Section \ref{sec:low-light driven training dataset}). 
Then, we present a novel attention network to recognize dark areas (Section \ref{sec:attention}).  
In Section \ref{sec:architecture}, we describe the proposed low-light enhancement method with detailed network architectures and loss functions.

\subsection{Low-light Driven Training Dataset} \label{sec:low-light driven training dataset}
To construct a large set of paired data, the deraining method in ~\cite{yang2017deep} synthesized raindrops.
The dehazing methods in~\cite{cai2016dehazenet, li2017reside} synthesized haze images using depth information through single image depth estimation. 
For low-light enhancement, methods in~\cite{Chongyi18, Shen17} synthesized low-light images. 
However, it is nontrivial for these methods to accurately describe illumination in real-world. 
Most conventional methods synthesized global illumination as shown in Fig.\ref{fig:illumination synthesis compare}(b). 
However, these methods are only suitable for under exposure images and cannot handle low-light images properly if dark areas are caused by natural illumination. 
Under the real-world environment, global and local illumination can exist simultaneously in the same scene. 
If we attempt to apply global illumination to real-world images, normally illuminated areas in the images can be easily saturated. 
To solve this problem, low-lightGAN~\cite{Kim_ICIP19} presented a quad-tree local illumination synthesis method, as shown in Fig.\ref{fig:illumination synthesis compare}(c). 

We propose a local illumination synthesis method based on superpixels, as shown in Fig.\ref{fig:illumination synthesis compare}(d). 
We apply a randomly different level of illumination to each superpixel and synthesize both the low-light and normal illumination areas. 
In addition, the proposed superpixel-based method can synthesize local illumination according to object boundaries, because superpixels describe object shapes. 
The local illumination synthesis using superpixel is formulated as follows:
\begin{equation}\label{local_illum}
     I_{local} = SLIC(I) \times L,~\text{for}~L=\{0.1, 0.2, ... 0.9, 1.0\},
\end{equation}
where the SLIC function~\cite{achanta2012slic} outputs superpixels of image $I$. 
In \eqref{local_illum}, $L$ denotes the illumination weight. 
If $L=1.0$, the original brightness is maintained. 
If $L=0.1$, the corresponding superpixel is considerably darkened. 
This synthetic data is used to train the dark-aware attention network, which is explained in the next section.

\begin{figure*}[t]
    \begin{minipage}[b]{1.0\linewidth}
        \centering
        \includegraphics[width=0.9\linewidth]{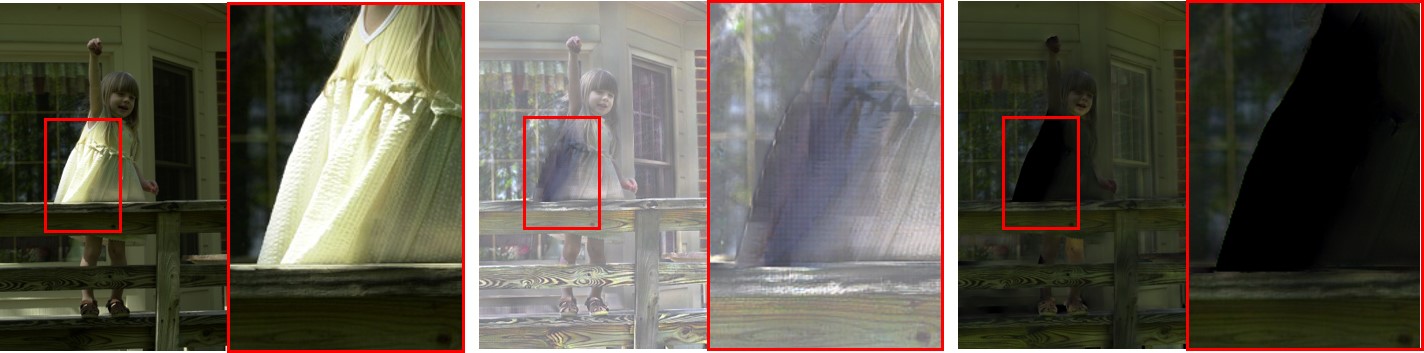}
    \end{minipage}
     \begin{minipage}[b]{0.32\linewidth}
            \centering
            {(a) }
    \end{minipage}
           \begin{minipage}[b]{0.32\linewidth}
            \centering
            {(b) }
    \end{minipage}
            \begin{minipage}[b]{0.32\linewidth}
            \centering
            {(c) }
    \end{minipage}
  \caption{\textbf{Proposed visual attention map}. (a) Input image, (b) Quad-tree-based method~\cite{Kim_ICIP19}, and (c) Our dark-aware attention method.
  }
 \label{fig:diff_quad_tree_superpixel}
\end{figure*}

\subsection{Dark-aware Visual Attention} \label{sec:attention}
The methods in \cite{Guo17, Xueyang, shen2009color} used existing attention mechanisms and produced illumination maps based on retinex theory. 
Lv and Lu \cite{lv2019attention} estimated attention maps based on max channels in original and low-light images.
CWAN~\cite{atoum2019color} created binary mask maps for the foreground color and used the maps for training. 
EnlightenGAN~\cite{jiang2019enlightengan} used illumination channels to estimate a self-regularized attention map. 

Unlike conventional attention-based low-light enhancement methods, our method can recognize visually dark areas rather than illumination areas based on retinex theory.
Thus, our method is a visual attention method (\ie dark-aware attention network), which focuses on dark areas for low-light image enhancement.
To train the proposed dark-aware attention network in a supervised manner, we synthesize the ground-truth attention map $I_{VA}$, as follows:
\begin{equation}\label{visual_attention_gt}
     I_{VA} = I - I_{local},
\end{equation}
where $I$ and $I_{local}$ in \eqref{visual_attention_gt} denote the original image and locally illuminated image, respectively. 
Local illumination and superpixels with various forms enable the proposed attention network to learn various types of bright and dark areas during the training. 
Fig.\ref{fig:diff_quad_tree_superpixel}(c) shows the estimated attention map of the proposed method.
Our attention map accurately differentiates between dark areas and bright regions, where bright regions are represented using small values (\ie red box) and dark areas are represented using large values. 
The conventional method~\cite{Kim_ICIP19} cannot accurately estimate visual attention maps with blocking artifacts, as shown in Fig.\ref{fig:diff_quad_tree_superpixel}(b).

\begin{figure*}[t]
    \begin{minipage}[b]{1.0\linewidth}
        \centering
        \includegraphics[width=0.9\linewidth]{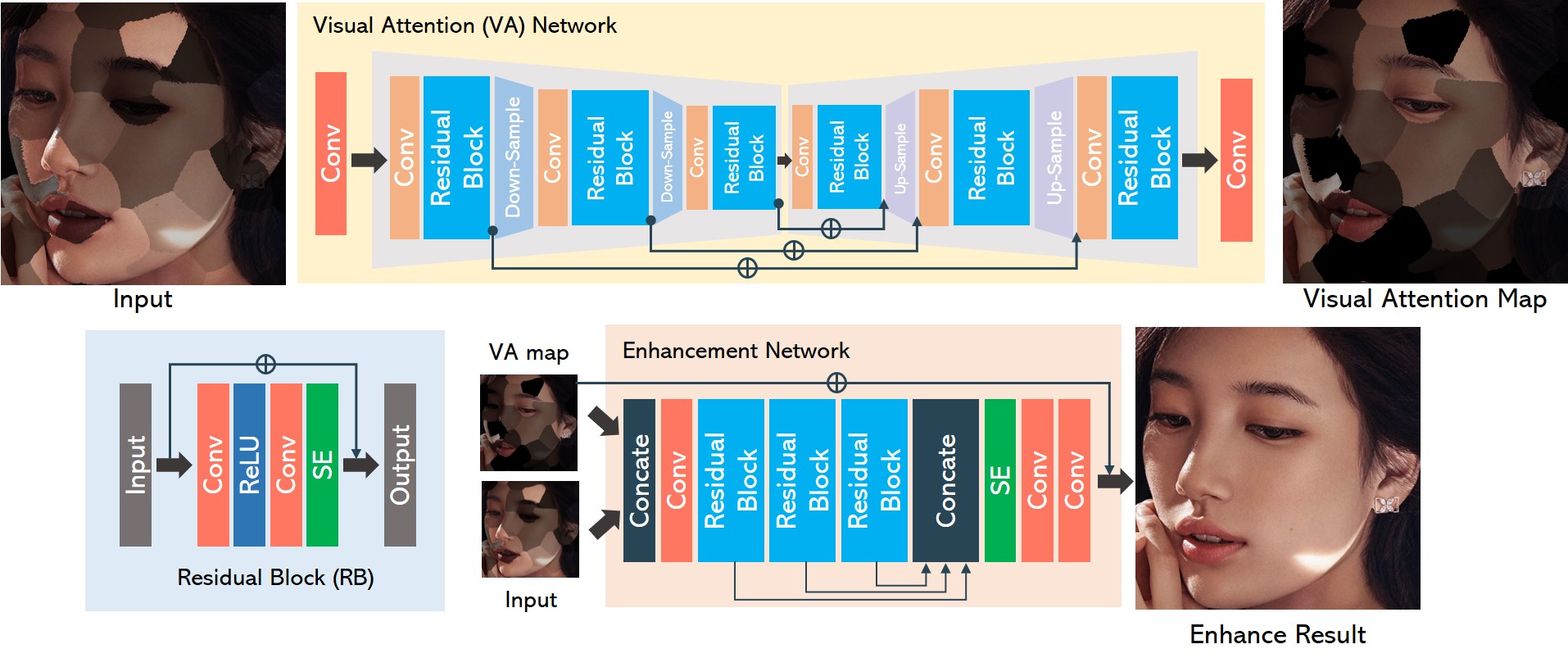}
    \end{minipage}
    \vspace{-6mm}
  \caption{\textbf{The proposed network architecture} consisting of visual attention network (VAN), enhancement network (EN), and residual block (RB). 
  }
 \label{fig:network-architecture}
\end{figure*}

\subsection{Dark Region-aware Low-light Enhancement Network} \label{sec:architecture}
The proposed low-light enhancement network consists of visual attention and enhancement networks, as shown in Fig.\ref{fig:network-architecture}.
The attention network produces the attention map that can recognize dark areas, whereas the enhancement network outputs low-light enhanced images.

\textbf{Visual Attention Network (VAN).} The proposed VAN adopts the U-Net~\cite{ronneberger2015u} structure (\ie encoder and decoder) as a backbone network. 
The first convolution layer has the kernel of $3 \times 3$ size with stride $1$.
The Encoder has three convolution layers, residual blocks, and two down-sample layers, as shown in Fig.\ref{fig:network-architecture}. 
Residual Blocks consist of convolution layers with the kernels of $1 \times 1$ size, ReLUs, and squeeze-and-excitation blocks~\cite{hu2018squeeze}, which have different dilation factors (\ie $3$, $2$, and $1$). 
The decoder has three convolution layers, residual blocks, and two up-sample layers. 
Residual Blocks use different dilation factors (\ie $1$, $2$, and $3$). 

We compute $l_2$ loss between the estimated attention map $VA(I_{local})$ and ground truth $I_{VAGT}$ at the pixel level:
\begin{equation}\label{attention_loss}
    \mathcal{L}_{a} = \left\|VA(I_{local}) - I_{VAGT}\right\|_2,
\end{equation}
where $I_{local}$ in \eqref{local_illum} denotes the low-light image with locally illuminated areas and the function $VA$ estimates visual attention maps.
We also calculate perceptual loss, which measures the similarity between $VA(I_{local})$ and $I_{VAGT}$ at the feature level:  
\begin{equation}\label{perceptual_loss}
    \mathcal{L}_{p} = \|\phi(VA\left(I_{local})+I_{local}\right) - \phi(I_{GT})\|_1,
\end{equation}
where $\phi$ is $16$-th feature map obtained by the pre-trained VGG-16 network~\cite{Chatfield14} and $I_{GT}$ denotes ground truth for the low-light enhanced image.
Then, the total loss for VAN is designed as follows. 
\begin{equation}\label{enhance_total_loss}
     \mathcal{L}_{VAN} = \lambda_{1}\mathcal{L}_{a} + \lambda_{2}\mathcal{L}_{p},
\end{equation}
where $\lambda_{1}$ and  $\lambda_{2}$ are weighting hyper-parameters. 

\textbf{Enhancement Network (EN).} The proposed EN enhances the brightness of low-light images using the estimated visual attention maps. 
The EN takes the concatenation of low-light image and visual attention map as an input. 
Similar to the VAN, all convolution layers have the kernel of $3 \times 3$ size with stride $1$.
Three residual blocks use different dilation factors from $3$ to $1$, while we concatenate all residual blocks to fuse the information. 

We compute $l_2$ loss between the low-light enhanced image $EN(I_{EN})$ and ground truth $I_{GT}$ at the pixel level:
\begin{equation}\label{enhance_loss}
    \mathcal{L}_{e} = \|EN(I_{EN}) - I_{GT}\|_2,
\end{equation}
where $I_{EN}=VAN(I_{local})+I_{local}$.
We also calculate perceptual loss, which measures the similarity between $EN(I_{EN})$ and $I_{GT}$ at the feature level:
\begin{equation}\label{enhance_perceptual_loss}
     \mathcal{L}_{ep} = \|\phi(EN(I_{EN})) - \phi(I_{GT})\|_1.
\end{equation}
The total variation loss aims to make output images spatially smooth:
\begin{equation}\label{tv_loss}
     \mathcal{L}_{tv} = \frac{1}{CHW} \|\bigtriangledown_{x} EN(I_{EN}) + \bigtriangledown_{y}EN(I_{EN})\|^2,
\end{equation}
where $\bigtriangledown_{x}$ and $\bigtriangledown_{y}$ differentiate the images via the $x$ and $y$ directions. $C$, $H$, and $W$ are the channel, height, and width of the enhanced image, respectively.
Then, the total loss for EN is designed as follows. 
\begin{equation}\label{HR_total_loss}
     \mathcal{L}_{EN} = \lambda_{1}\mathcal{L}_{e} + \lambda_{2}\mathcal{L}_{ep}  + \lambda_{3}\mathcal{L}_{tv}.
\end{equation}


\section{Experiments}
     \vspace{-2mm}

\begin{figure*}[t]
    \begin{minipage}[b]{1.0\linewidth}
        \centering
        \includegraphics[width=0.9\linewidth]{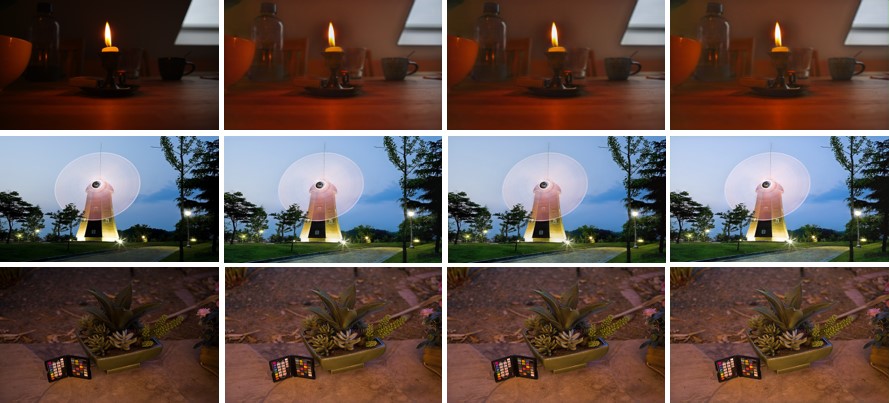}
    \end{minipage}
         \begin{minipage}[b]{0.22\linewidth}
            \centering
            {\footnotesize{(a) Input}}
    \end{minipage}
    \begin{minipage}[b]{0.28\linewidth}
            \centering
            {\footnotesize{(b) Input+visual attention map}}
    \end{minipage}
    \begin{minipage}[b]{0.24\linewidth}
            \centering
            {\footnotesize{(c) Without refinement}}
    \end{minipage}
    \begin{minipage}[b]{0.2\linewidth}
            \centering
            {\footnotesize{(d) With refinement}}
    \end{minipage}
    \vspace{-3mm}
  \caption{\textbf{Ablation Study 1: the proposed adversarial learning refinement.}
  }
 \label{fig:GAN}
       \vspace{-3mm}
\end{figure*}
\begin{figure*}[t]
    \begin{minipage}[b]{1.0\linewidth}
        \centering
        \includegraphics[width=0.9\linewidth]{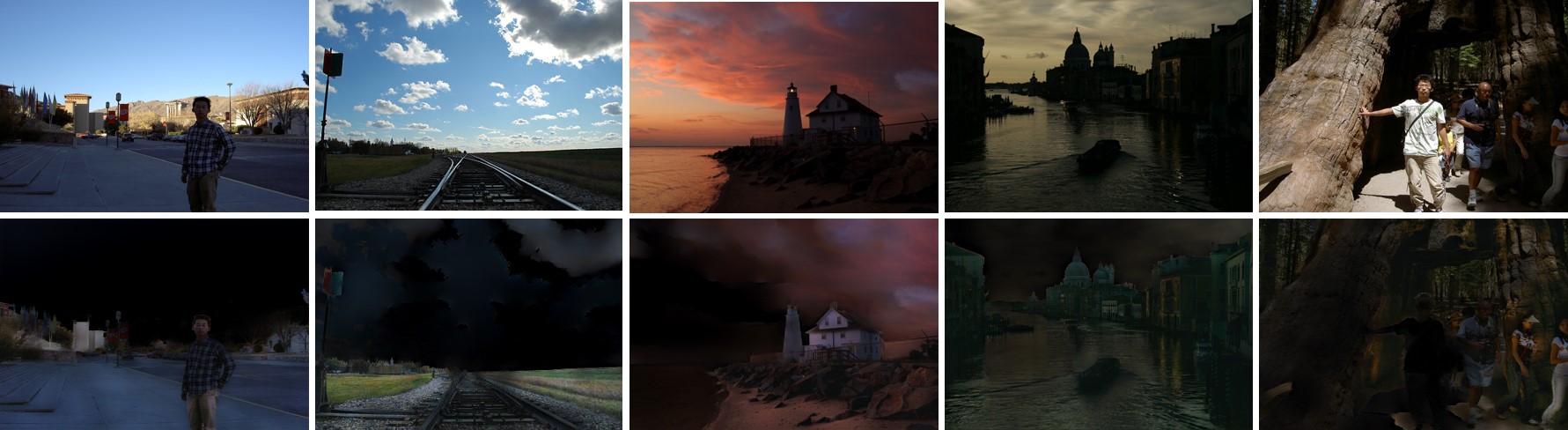}
    \end{minipage}
       \vspace{-6mm}
  \caption{\textbf{Ablation Study 2: the proposed visual attention map}. First row: input images, Second row: estimated visual attention maps.}
 \label{fig:visual_attention_map}
        \vspace{-4mm}
\end{figure*}

For training, we synthesized locally illuminated images using the DIV2K~\cite{timofte2017ntire} and Flickr2K \cite{timofte2017ntire} datasets, as explained Section \ref{sec:low-light driven training dataset}.  
The proposed visual attention network was trained with learning rate of $1e-5$ for 130 epochs, whereas the proposed enhancement network was trained with learning rate of $1e-5$ for 40 epochs.
We trained the whole network for $24$ hours on NVIDIA GeForce GTX TITAN Xp GPUs. 
Hyper-parameters $\lambda_1$ and $\lambda_2$ in \eqref{enhance_total_loss} were set to $0.5$ and $1$, respectively
Hyper-parameters $\lambda_1$, $\lambda_2$, and $\lambda_3$ in \eqref{HR_total_loss} were set to $1$, $5$, and $1$, respectively.
We randomly cropped approximately $3500$ images with $2K$ resolutions into $240 \times 240$ patches. 
We compared our method (\textbf{DALE}) with state-of-the-art methods including  NPE~\cite{wang2013naturalness}, LIME~\cite{Guo17}, MEF~\cite{ma2015perceptual}, and DICM~\cite{lee2012contrast}.

\subsection{Ablation Study}
     \vspace{-2mm}
\noindent\textbf{Refinement via adversarial learning.} 
The proposed visual attention intensively enhances the brightness of dark areas, while it preserves the brightness of other regions. 
Thus, if the visual attention map is accurately estimated, the following enhancement network has little effect on the performance of low-light enhancement and parameters of the network are hardly updated. 
To solve this problem, we trained the network in an adversarial manner (\ie DALEGAN).
We used the enhancement network as a generative network and employed PatchGAN~\cite{isola2017image,miyato2018spectral} as a discriminator network.
Fig.\ref{fig:GAN} shows that the enhancement network with adversarial learning further enhanced the brightness of low-light region.
\begin{table*}[t]
\caption{\textbf{Quantitative comparison with state-of-the-art low-light enhancement methods using NIQE and BRISQUE}. Red and blue numbers denote the best and second best results, respectively.}
\setlength{\tabcolsep}{20pt}
\begin{adjustbox}{max width=\textwidth}
\begin{tabular}{lcccc}
\hline\hline
{NIQE / BRISQUE}                                      & DICM      & LIME      & MEF       & NPE       \\ \hline
LIME~\cite{Guo17}                                     & 3.63 / 26.8 & 4.35 / 22.3 & 3.83 / 24.1 & 3.84 / 26.1 \\
BIMEF~\cite{ying2017bio}                              & {\color{blue}{3.38}} / 26.8 & {\color{blue}{3.55}} / 23.2 & {\color{blue}{3.13}} / {\color{red}{19.3}} & 3.5 / 24.5  \\
RetinexNet~\cite{Chen2018Retinex}                     & 4.31 / 26.7 & 4.91 / 26.1 & 4.90 / 26.0 & 4.07 / 26.9 \\
EnlightenGAN~\cite{jiang2019enlightengan}             & {\color{red}{3.05}} / 26.3 & {\color{red}{3.37}} / {\color{red}{20.6}} & {\color{red}{2.89}} / 23.6 & {\color{blue}{3.34}} / 27.3 \\
DALE                                                  & 3.78 / {\color{red}{21.4}} & 4.33 / \color{blue}{22.2} & 4.01 / {\color{blue}{22.8}} & 3.38 / {\color{blue}{22.1}} \\ 
DALEGAN                                               & 3.61 / \color{blue}{22.2} & 4.16 / 23.5 & 3.80 / 23.9 & {\color{red}{3.31}} / {\color{red}{19.6}} \\ \hline\hline
\end{tabular}
\label{table:niqe}
\end{adjustbox}
\vspace{-3mm}
\end{table*}

\begin{table*}[t]
\caption{\textbf{Quantitative comparison with state-of-the-art low-light enhancement methods using LOE}. Red and blue numbers denote the best and second best results, respectively.}
\setlength{\tabcolsep}{20pt}
\begin{adjustbox}{max width=\textwidth}
\begin{tabular}{lcccc}
\hline\hline
{LOE}                                       & DICM      & LIME      & MEF       & NPE       \\ \hline
LIME~\cite{Guo17}                           & 1260.8    & 1323.8    & 1079.4    & 1119.6    \\
BIMEF~\cite{ying2017bio}                    & \color{red}{351.82}    & \color{red}{478.57}    & \color{red}{325.86}    & \color{red}{308.12}    \\
RetinexNet~\cite{Chen2018Retinex}           & 1565.8    & 1882.5    & 1777.4    & 1224.5    \\
EnlightenGAN~\cite{jiang2019enlightengan}   & 1318.8    & 1361.5    & 1141.9    & 1346.2    \\
DALE                                        & \color{blue}{888.7}     & \color{blue}{810}       & \color{blue}{829.2}     & \color{blue}{678.7}     \\ 
DALEGAN                                     & 920.91    & 849.6     & 892.4     & 714.6     \\ \hline\hline 
\end{tabular}
\end{adjustbox}
\label{table:loe}
\vspace{-5mm}
\end{table*}

\noindent\textbf{Dark-aware visual attention.} 
Fig.\ref{fig:visual_attention_map} shows examples of estimated visual attention maps.
The proposed visual attention method accurately recognized dark areas in low-light images.
The method did not focus on bright areas such as sky regions, because the regions appear clearly and there exist sufficient brightness.
In the last column of Fig.\ref{fig:visual_attention_map}, the visual attention method produced very small values for a tree or left person, because they received direct sunlight.
In contrast, the method produced large values for a right person, because he was in dark areas.
\subsection{Comparison with Other Methods}
We quantitatively evaluated several methods in terms of lightness order error (LOE)~\cite{wang2013naturalness}, naturalness image quality evaluator (NIQE) and blind/referenceless image spatial quality evaluator (BRISQUE)~\cite{mittal2012no} \ie no-reference image quality evaluation metric in~\cite{mittal2013making}.
Tables \ref{table:niqe} and \ref{table:loe} show the low-light enhancement results for four standard benchmark datasets, which contain real-world low-light images.
The proposed DALE and EnlightenGAN showed state-of-the-art performance in terms of NIQE and BRISQUE, where overall image quality has been improved.
In terms LOE, BIMEF produced the best results.
Please note that LOE evaluates the degree of light distortion. 
However, it produces good scores (\ie small values), even though the brightness of images is not much improved.

\begin{figure*}[t]
    \begin{minipage}[b]{1.0\linewidth}
        \centering
        \includegraphics[width=0.77\linewidth]{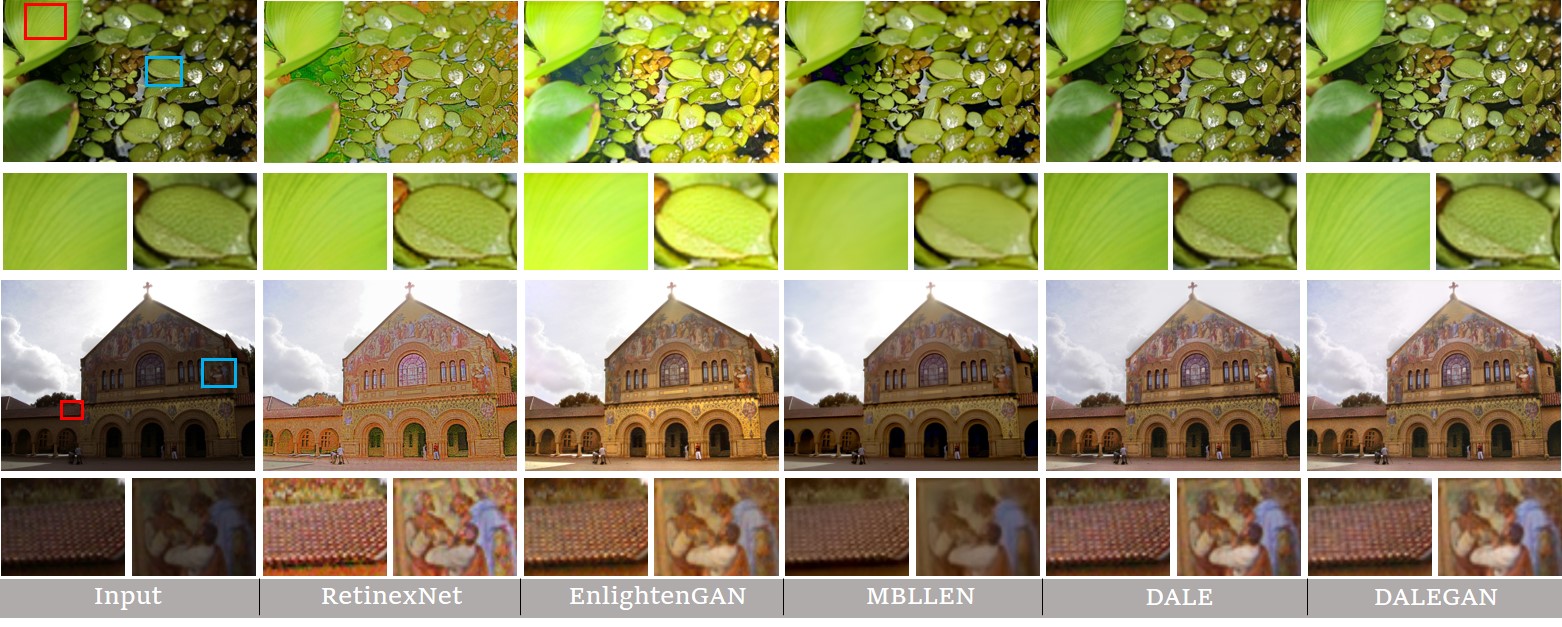}
    \end{minipage}
     \begin{minipage}[b]{1.0\linewidth}
            \centering
            {(a) Qualitative comparison on saturation and details}
    \end{minipage}
    \begin{minipage}[b]{1.0\linewidth}
        \centering
        \includegraphics[width=0.77\linewidth]{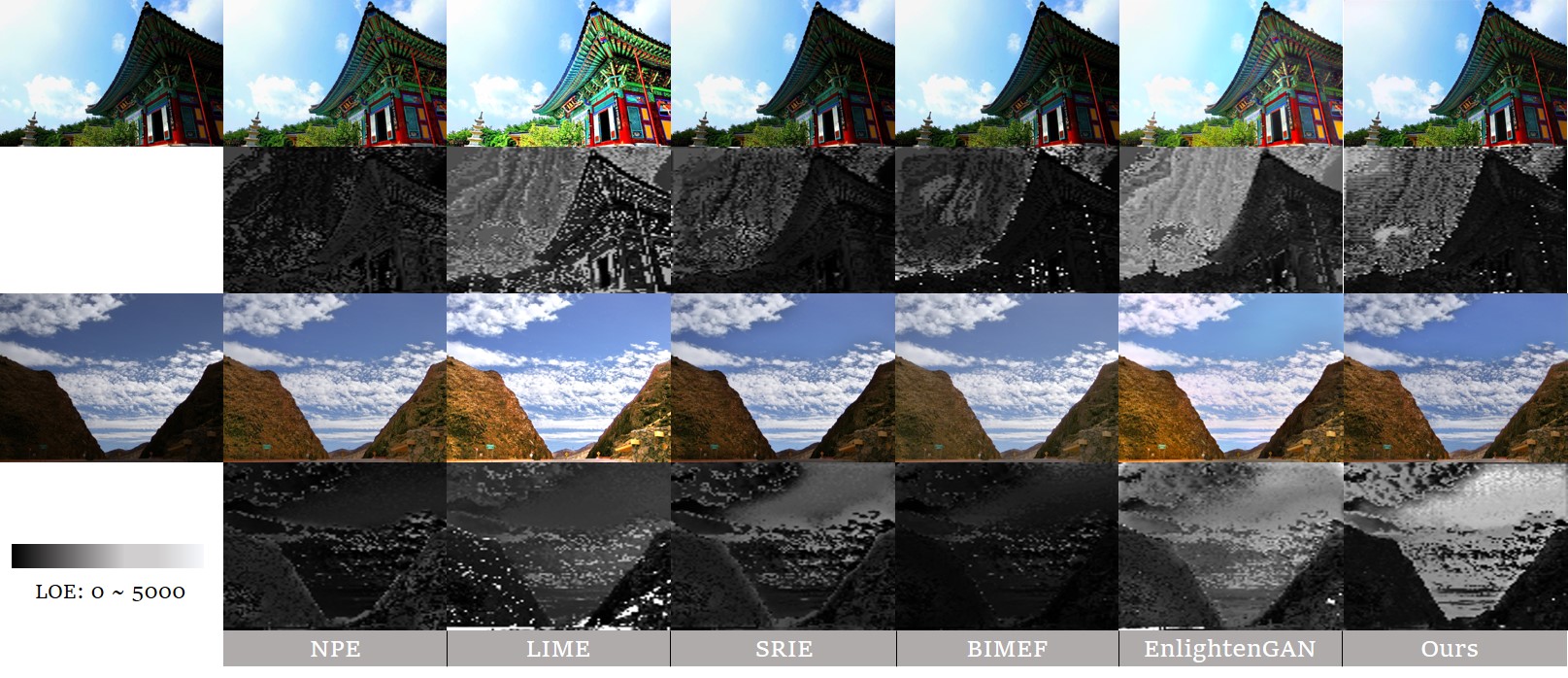}
    \end{minipage}
     \begin{minipage}[b]{1.0\linewidth}
            \centering
            {(b) Qualitative comparison on lightness distortion}
    \end{minipage}
    \begin{minipage}[b]{1.0\linewidth}
        \centering
        \includegraphics[width=0.77\linewidth]{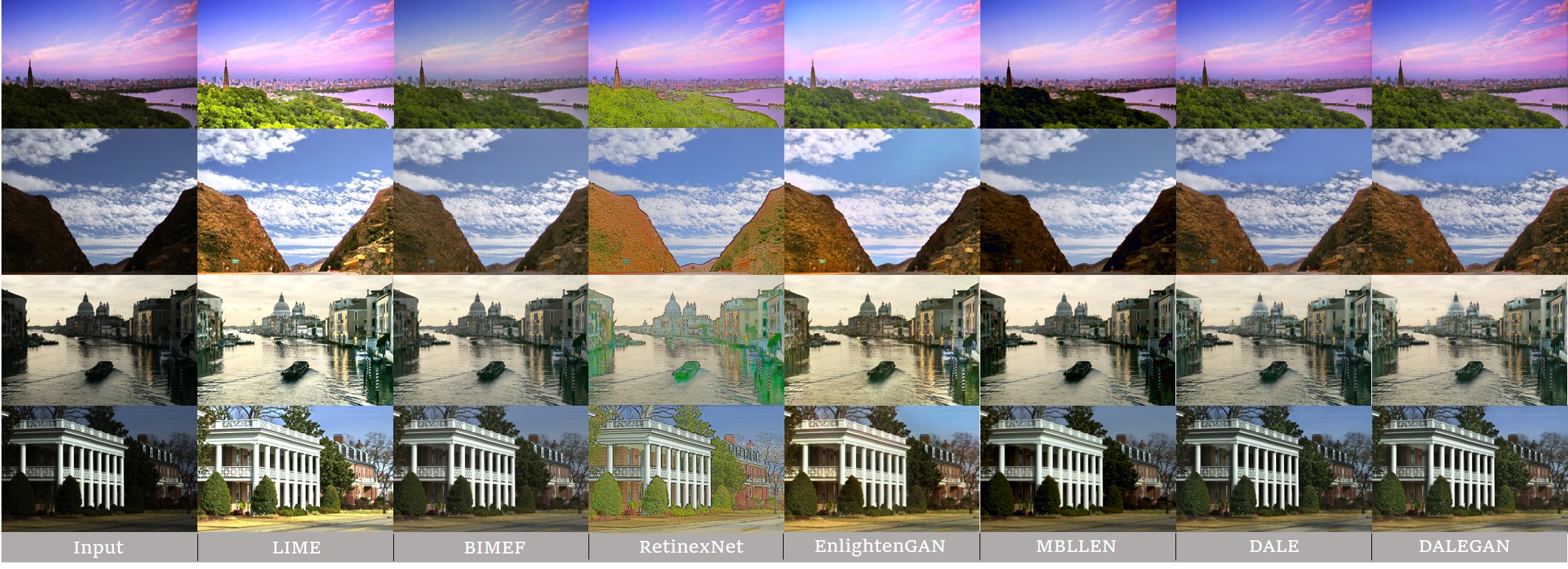}
    \end{minipage}
     \begin{minipage}[b]{1.0\linewidth}
            \centering
            {(c) Qualitative comparison with other methods}
    \end{minipage}
    \vspace{-7mm}
  \caption{\textbf{Qualitative comparison with state-of-the-art-methods}. (b) first and third rows: input images, second and fourth rows: estimated LOE maps.}
 \label{fig:qualitative}
     \vspace{-3mm}
\end{figure*}

Thus, BIMEF still has visually low-light regions, as shown in the second row of Fig.\ref{fig:qualitative}(b).
In contrast, LIME, EnlightenGAN, and the proposed DALE produced qualitatively better low-light enhance results than BIMEF, although they have very high values of LOE.

Fig.\ref{fig:qualitative}(a) qualitatively compared saturation and details of low-light enhanced images. 
Fig.\ref{fig:qualitative}(b) qualitatively compared lightness distortion using LOE maps in low-light enhanced images.
As shown in the third row of Fig.\ref{fig:qualitative}(c), LIME over-improved the brightness of images, which results in saturation problmes. 
BIMEF and MBLLEN produced clear images but lacked the brightness. 
EnlightenGAN produced good image enhancement results.
However, some colors distorted. 
In contrast these methods, the proposed DALE qualitatively outperforms other state-of-the-art methods. 
       \vspace{-3mm}

\section{Conclusion}
     \vspace{-3mm}
In this paper, we present a novel low-light enhancement method (DALE) based on a new visual attention network that can recognize dark regions.
We synthesize different local illumination for each super-pixel and accurately estimate the brightness of low-light images using synthetic training images.
Experimental results demonstrate that our method outperforms state-of-the-art methods. 
\newline

\noindent\textbf{Acknowledgments:} This research was supported by the the Seoul R\&BD Program (CY190032) and the ITRC program (IITP-2020-2018-0-01799).

\bibliography{egbib}
\end{document}